\documentclass[aps,prd,a4paper,twocolumn,showpacs,showkeys,preprintnumbers,amsmath,amssymb,nofootinbib,usenames,dvipsnames]{revtex4-2}
\usepackage{graphicx}
\usepackage{bm}
\usepackage{dcolumn}
\usepackage{xcolor}
\usepackage{multirow}

\usepackage[titletoc]{appendix}

\usepackage[top=2cm,bottom=3cm,left=2.1cm,right=1.0cm]{geometry}

\usepackage{mathrsfs} 
\usepackage{txfonts} 

\newcommand{\bea}{\begin{eqnarray}}
\newcommand{\eea}{\end{eqnarray}}

\def\beq{\begin{equation}}
\def\eeq{\end{equation}}

\newcommand{\rd}{\mathrm{d}}  

\makeatletter
\def\p@subsection{}
\def\p@subsubsection{}
\makeatother

\def\al{\alpha}
\def\be{\beta}

\def\th{\theta}

\def\si{\sigma}

\def\ta{\tau}

\begin{document}

\title{Neglected solutions in quadratic gravity}

\author{Breno L. Giacchini}
\email{breno.giacchini@matfyz.cuni.cz}
\affiliation{
{\small Institute of Theoretical Physics, Faculty of Mathematics and Physics, Charles University, V Hole{\v s}ovi{\v c}k{\'a}ch 2, 180 00 Prague 8, Czech Republic}
}

\author{Ivan Kol\'a\v{r}}
\email{ivan.kolar@matfyz.cuni.cz}
\affiliation{
{\small Institute of Theoretical Physics, Faculty of Mathematics and Physics, Charles University, V Hole{\v s}ovi{\v c}k{\'a}ch 2, 180 00 Prague 8, Czech Republic}
}

\date{\small\today}

\begin{abstract} 
\noindent
We report on several previously overlooked families of static spherically symmetric solutions in quadratic gravity. Our main result concerns the existence of solutions whose leading exponents depend on the ratio ${\omega=\alpha/(3\beta)}$ of the four-derivative couplings. We demonstrate that the space of models with ${\omega >1}$ contains a dense set that admits non-Frobenius solutions ${(s_*, 2 - 3 s_*)_0}$ (in standard Schwarzschild coordinates), with certain rational numbers $s_*(\omega)$. These solutions correspond to a singular core at ${\bar{r}=0}$. Another related non-Frobenius family, $(s_*, 2 - 3 s_*)_\infty$, exists for a dense set of models with ${1/4 < \omega < 1}$, describing a singular boundary at ${\bar{r}\to\infty}$. Both families are uncovered by recasting the metric into special coordinates in which the solutions become Frobenius. Additionally, for models with real ratios ${\omega\neq 1}$ we identify six novel families of non-Frobenius solutions around points ${\bar{r}=\bar{r}_0} \neq 0$, describing horizons and wormhole throats. Finally, we re-derive and summarize all known families of solutions in modified as well as in the standard Schwarzschild coordinates.
\end{abstract}

\maketitle
\noindent


\section{Introduction}

Quadratic gravity has a prominent role in the formulation of a field-theory approach to quantum gravity. In being renormalizable it outperforms general relativity, but falls short when it comes to unitarity~\cite{Stelle77}. It is relatively simple if compared to other renormalizable models (see, e.g.,~\cite{hdgnl}), and considerable progress has been made to understand its quantum and classical properties, especially regarding restoration of unitarity~\cite{unitarity} and stability of solutions~\cite{stability}.
The first classification of static, spherically symmetric solutions of a general model of quadratic gravity, defined by the action
\beq
S = \int \rd^4 x\, \sqrt{-g}\, \left( 
\gamma \, R +\beta\,R^2  - \alpha\, C_{\mu\nu\alpha\beta}\, C^{\mu\nu\alpha\beta}	\right)  
\label{action-4der}
\eeq
with arbitrary couplings $\alpha$, $\beta$ and $\gamma$, 
was the 1978 paper by K.~Stelle~\cite{Stelle78}. That work considered vacuum solutions that can be written in the form
\beq
\label{metric-Standard}
\rd s^2 = - B(\bar{r}) \rd t^2 + \frac{ \rd \bar{r}^2 }{A(\bar{r})} + \bar{r}^2 \left( \rd \th^2 + \sin^2 \th \, \rd \phi^2 \right) 
,
\eeq
where the functions $A(\bar{r})$ and $B(\bar{r})$ admit a representation in Frobenius series around the origin $\bar{r} = \bar{r}_0 = 0$:
\beq
\begin{split}
\label{FrobeniusAB}
A(\bar{r})   =  & \, (\bar{r} - \bar{r}_0)^s \sum_{n=0}^\infty a_{n} \, (\bar{r} - \bar{r}_0)^n ,
\\
B(\bar{r})  = & \, b_{0} (\bar{r} - \bar{r}_0)^t \Bigg[ 1 + \sum_{n=1}^\infty b_{n} \, (\bar{r} - \bar{r}_0)^n  \Bigg]  , \quad \,\, a_0,b_0 \neq 0,
\end{split}
\eeq
with the exponents $s$ and $t$ governing the leading behaviour of the solution. 
By solving the indicial equations (\textit{i.e.}, the field equations expanded to the lowest order in $\bar{r}$), it was found that the only solutions ``for general values of the parameters $\alpha$ and $\beta$'' are $(s,t) \in \lbrace (0,0), \,  (-1,-1), \, (-2,2) \rbrace$~\cite{Stelle78}.

That work was revisited in 2015, when the previous analysis was expanded to solutions~\eqref{FrobeniusAB} expanded around other points $\bar{r} = \bar{r}_0 \neq 0$~\cite{Stelle15PRL,Stelle15PRD} (see also~\cite{Holdom:2002} for earlier related results). This led to the discovery of new black hole and wormhole solutions. Also, some solutions in non-Frobenius form were found, for which the series in~\eqref{FrobeniusAB} follow half-integer steps; they represent unusual horizons and non-symmetric wormholes. 
Regarding the solutions around $\bar{r} = 0$, the study verified and extended the results of~\cite{Stelle78} and, in a footnote, it mentioned three additional families of solutions that exist for particular quadratic gravity models~\cite{Stelle15PRD}: two for the case $\alpha =0$ and one for models with special values of the combination
\beq
\frac{\alpha}{3\beta} \equiv \omega,
\eeq
namely,
\beq
\label{solcond}
\frac{2-t}{3} = s \in \mathbb{Z}^- , \qquad  \omega  = \frac{(s^2 - 2s +2)^2}{s^4} .
\eeq
(Note that, in our notation, $s$ is defined with opposite sign.)

This last family of solutions is the central topic of the present Letter. We argue that, although only for $s \in \mathbb{Z}^-$ the functions $A(\bar{r})$ and $B(\bar{r})$ have an expansion in Frobenius series, there are non-Frobenius solutions for other negative values of $s$. Specifically, we prove that the set of models that admit such solutions is dense in the space of quadratic gravity models with $\omega > 1$. Thus, there seems to exist a consistent family of solutions, with one free physical parameter and indicial structure $(s,t)=(s_*, 2 -3s_*)$ such that $s_*(\omega)$ is implicitly defined by
\beq
\label{star_def}
\omega = \frac{(s_*^2 - 2s_* +2)^2}{s_*^4}  , 
\eeq
restricted to $s_* < 0$.
We shall denote this family of solutions by $(s_*, 2-3s_*)_0$, with the subscript indicating that they are expansions around $\bar{r}=0$.

Most of our discussion uses modified Schwarzschild coordinates and, for the sake of completeness, we also scanned the possible solutions in Frobenius form in those coordinates. This led to the discovery of additional families of solutions that were overlooked in the literature. 
The paper is organised as follows: 
first, we prove the main result regarding the existence of the solution family $(s_*, 2-3s_*)_0$. 
Then, we
discuss the other new families, 
followed by a summary of all the known vacuum solutions.

\section{The family $(s_*, 2-3s_*)_0$}
\label{Sec2}

The solutions in this family  
are not usually in the form of Frobenius series in $\bar{r}$ in standard spherically symmetric coordinates~\eqref{metric-Standard}. Therefore, we shall use other coordinates that provide a simpler description. The inspiration comes from the use of Kundt coordinates in Einstein--Weyl gravity~\cite{Podolsky:2018pfe,Svarc:2018coe,Podolsky:2019gro} and modified Schwarzschild coordinates in six-derivative gravity~\cite{Giacchini:2025gzw} as, in both cases, solutions that are not Frobenius series in Schwarzschild coordinates take the form of conventional power series. 
We start with modified Schwarzschild coordinates and the metric ansatz
\beq
\label{metric}
\rd s^2 = - H(r) \rd t^2 + \frac{\rd r^2}{H(r)} + F^2(r) \left( \rd \th^2 + \sin^2 \th \rd \phi^2 \right) 
,
\eeq
which yields autonomous field equations, \textit{i.e.}, the dependence on $r$ is through the functions $F$, $H$ and their derivatives.  
The relation between a solution written in this form and~\eqref{metric-Standard} is via
\beq
\label{Fr}
\bar{r} = F(r),
\eeq
which gives $A(\bar{r}) = H(r) F^{\prime 2}(r)$ and  $B(\bar{r}) = H(r)$,
with $r=r(\bar{r})$ obtained by inverting~\eqref{Fr}. The metric~\eqref{metric} admits a 2-parameter residual gauge freedom consisting of a shift of the coordinate $r$, and a simultaneous re-scaling of $r$ and $t$; on the other hand, the metric~\eqref{metric-Standard} only admits the time re-scaling.

If a solution has the leading behaviour $F(r) \sim (r-r_0)^\sigma$ and $H(r) \sim (r-r_0)^\tau$ with $\sigma > 0$ around a certain point $r=r_0$, from~\eqref{Fr} it follows that $r_0$ corresponds to the origin ($\bar{r}=0$) in Schwarzschild coordinates, and it is not difficult to verify that $A(\bar{r}) \sim \bar{r}^s$ and $B(\bar{r}) \sim \bar{r}^t$ with
\beq
\label{st1}
s = \frac{2\si + \tau - 2}{\si} , \qquad t = \frac{\tau}{\si} .
\eeq
In this spirit, in the modified coordinates~\eqref{metric}, the solutions that we seek are characterised by
\beq
\label{def2}
\sigma = \frac{3}{2}-\tau , 
\quad \,\,\,\,
\omega =\frac{(2 \tau ^2-6 \tau +5)^2}{4 (\tau -1)^4} , \quad \,\,\,\, 1 < \tau < \frac{3}{2} .
\eeq
These relations uniquely define $\tau$ as a function of $\omega$,
\beq
\label{TauInv}
\tau_+ (\omega) = \frac{1}{2} \left( 2 - \frac{1}{\sqrt{\omega }-1} +  
\sqrt{\frac{2 \sqrt{\omega }-1}{\big(\sqrt{\omega }-1\big)^2}}\right) , \quad \,\, \omega  > 1,
\eeq
leading to the indicial structure $\lbrace  \sigma ,\tau \rbrace =  \left\lbrace  3/2 - \tau_+(\omega) , \tau_+(\omega) \right\rbrace $.

From now on, let us assume that $\tau$ is rational, that is, $\tau \in \left( 1,3/2\right) \cap \mathbb{Q}$. Consequently, there exist coprime naturals $p$ and $q$ satisfying $2 p/3<q<p$ 
such that
\beq
\tau = \frac{p}{q}.
\eeq
In terms of $p$ and $q$, Eq.~\eqref{def2} becomes
\beq
\label{ompq}
\sigma = \frac{3}{2}-\frac{p}{q} , \quad \omega = \frac{(2 p^2-6 p q+5 q^2)^2}{4 (p-q)^4} , \quad \frac{2 p}{3}<q<p .
\eeq
In this parameterisation, let us also assume that the functions $F(r)$ and $H(r)$ are Puiseux series with $1/q$ steps:
\beq
\label{Series-FH}
\begin{split}
& F(r) = \Delta^{\sigma} 
\sum_{n=0}^\infty  f_n \Delta^{\frac{n}{q}}  , \qquad 
H(r) = \Delta^{\tau} 
\sum_{n=0}^\infty h_n \Delta^{\frac{n}{q}} ,  
\\
& f_0,h_0\neq 0,  \qquad \Delta \equiv r-r_0.
\end{split}
\eeq
The above functions become standard Frobenius series (with unit steps) upon the definition
$$
u = \Delta^\frac{1}{q}.
$$
With this change of variables the metric can be cast in the form
\beq
\label{metric-U}
\rd s^2 = - \mathcal{H}(u) \rd t^2 + \frac{q^2 u^{2(q-1)}}{\mathcal{H}(u)} \rd u^2 + \mathcal{F}^2(u) \left( \rd \th^2 + \sin^2 \th \rd \phi^2 \right) ,
\eeq
with
\beq
\mathcal{F}(u) = u^{\frac{3}{2}q-p} \sum_{n=0}^\infty  f_n u^{n}  , \quad \mathcal{H}(u) = u^{p} \sum_{n=0}^\infty h_n u^{n} ,  \quad f_0,h_0\neq 0 .
\nonumber
\eeq
As the metric involves $\mathcal{F}^2(u) \sim u^{3q-2p}$, only integer exponents appear when the field equations are expanded around $u=0$.

Assuming~\eqref{ompq} and 
expanding the field equations in powers of $u$ it follows that the leading term is of order $u^{2 p-4 q+1}$:
\beq
\label{expeom}
\begin{split}
& \mathcal{E}^t{}_t = u^{2 p-4 q+1} \sum_{n=0}^\infty ( \mathcal{E}^t{}_t )_n u^n = 0 , 
\\
& \mathcal{E}^u{}_u = u^{2 p-4 q+1} \sum_{n=0}^\infty ( \mathcal{E}^u{}_u )_n u^n = 0 .
\end{split}
\eeq
(Note that once~\eqref{expeom} is solved, the generalised Bianchi identity $\nabla^\nu \mathcal{E}_{\mu\nu} = 0$ gives $\mathcal{E}^\theta{}_\theta = \mathcal{E}^\phi{}_\phi = 0$ as well.)
At the lowest order, it is possible to verify that $( \mathcal{E}^t{}_t )_0 = 0$ and $( \mathcal{E}^u{}_u )_0 = 0$ can be solved for the coefficients $f_1$ and $h_1$, leaving $f_0$ and $h_0$ free. More generally, at any given order $u^{2 p-4 q+N+1}$ with $N\in\mathbb{N}$, the equations $( \mathcal{E}^t{}_t )_N = 0$ and $( \mathcal{E}^u{}_u )_N = 0$ yield, respectively,
\beq
\label{E_N}
\begin{split}
\mathfrak{f}_1(p,q,N+1) h_0 f_{N+1} + \mathfrak{h}_1(p,q,N+1) f_0 h_{N+1} + \Phi_{1,N} & = 0,
\\
\mathfrak{f}_2(p,q,N+1) h_0 f_{N+1} + \mathfrak{h}_2(p,q,N+1) f_0 h_{N+1} + \Phi_{2,N} &= 0,
\end{split}
\eeq
where $\Phi_{1,N}$ and $\Phi_{2,N}$ depend on the coefficients $f_n$ and $h_n$ with $n=0,1,\ldots,N$,
and the explicit formulas for $\mathfrak{f}_{1,2}$ and $\mathfrak{h}_{1,2}$ are given in Appendix~\ref{SM}.
Hence, the system~\eqref{E_N} can be solved for $f_{N+1}$ and $h_{N+1}$,
\beq
\begin{split}
f_{N+1} = & \, - \frac{ \mathfrak{h}_2(p,q,N+1) \, \Phi_{1,N} - \mathfrak{h}_1(p,q,N+1) \, \Phi_{2,N} }{h_0 \, \mathfrak{d}(p,q,N+1) } , 
\\
h_{N+1} = & \, \frac{ \mathfrak{f}_2(p,q,N+1) \, \Phi_{1,N} - \mathfrak{f}_1(p,q,N+1) \, \Phi_{2,N} }{f_0 \, \mathfrak{d}(p,q,N+1) } , 
\\
\mathfrak{d} \equiv & \,\mathfrak{f}_1 \mathfrak{h}_2 - \mathfrak{f}_2 \mathfrak{h}_1 ,
\end{split}
\eeq
provided that the determinant $\mathfrak{d}(p,q,N+1) \neq 0$.\footnote{
If $\mathfrak{d} = 0$ for a certain $N$, this would mean that the system has no solution at that order or that there is a new free parameter $f_{N+1}$ or $h_{N+1}$, depending on the exact form of $\mathfrak{f}_{1,2}$, $\mathfrak{h}_{1,2}$,  $\Phi_{1,N}$ and $\Phi_{2,N}$. In the former situation, it might still be possible to solve for one of the otherwise free parameters $f_0$ or $h_0$.
Thus, $\mathfrak{d} = 0$ could indicate that there is no solution for those values of $p$ and $q$ or that the solution has a different structure of free parameters.}
For each pair $(p,q)$, $\mathfrak{d} = 0$ leads to a biquadratic Diophantine equation for $N$,
\bea
\label{det=0}
& & 2\big[N^4+9 (p-q)^2 (p-2 q)^2 \big] (2 p^2-6 p q+5 q^2)  - N^2 (28 p^4
\nonumber
\\
& & -156 p^3 q+334 p^2 q^2-324 p q^3+121 q^4) = 0.
\eea
Although we have no proof that~\eqref{det=0} does not have solutions, by using numerical methods 
we did not find any solution for $p$ smaller than 5000. (Recall that, for each value of $p$, $q$ can only assume integer values within the range $2p/3 < q< p$.) This suggests that only for very special values of $p$ and $q$ (if any) it could happen that $\mathfrak{d} = 0$ has a solution. The reasoning above guarantees that in most cases the field equations can be solved order by order, 
with no additional free parameters appearing
at higher orders.

%
%
%
%

Regarding the number of free parameters, since Eqs.~\eqref{E_N} with $N=0$ are solved for $f_1$ and $h_1$, and subsequently for $f_2$, $h_2$ and so on, we conclude that the solutions in the form~\eqref{metric-U} have two free parameters ($f_0$ and $h_0$). Returning to the coordinates~\eqref{metric}, there is the additional (non-physical) free parameter $r_0$. Due to the residual gauge freedom, only one parameter is physical. 

As an explicit example, consider the models with $\omega = 841/81$, that have a solution with $p=13$, $q=10$ and the indicial structure $\lbrace 1/5 ,13/10 \rbrace$ in modified Schwarzschild coordinates --- or $(-3/2,13/2)_0$ in the standard ones. Setting $r_0 = 0$, its first few terms read
\begin{align}
F(r)  = &\, f_0 r^{1/5} + \tfrac{800}{5759 f_0 h_0} r^{1/2} - \tfrac{3950775643510000}{780734156825259 f_0^3 h_0^2}r^{4/5} 
\nonumber
\\
& +\tfrac{47200 \gamma  f_0}{57477 \beta  h_0}  r^{9/10} + O \big( r^{11/10} \big),
\nonumber
\\
H(r)  = &\, h_0 r^{13/10} + \tfrac{4000}{9303 f_0^2} r^{8/5} + \tfrac{558190255705000}{60056473601943 f_0^4 h_0}  r^{19/10} 
\nonumber
\\
& -\tfrac{113950 \gamma}{57477 \beta } r^2 + O\big(r^{11/5}\big).
\label{example}
\end{align}
Note that although $r^{1/10}$ increments were assumed, many coefficients actually vanish. This seems to be a feature of the solutions in this family, with the effect that solutions with similar $\tau$ could have similar steps. For example, the pattern of exponents of the solution with $\tau=25/19 \approx 1.32$ (for the models with $\omega = 42025/5184$) looks similar to that above with $\tau=1.3$, despite their very different base steps.

All solutions in this family are such that $R \neq 0$ and they contain curvature singularities. For instance,
\beq
\begin{split}
& R  \underset{u \to 0}{=}  -\tfrac{3 h_0 (2 p^2-6 p q+5 q^2)}{2 q^2} u^{p-2 q} + \ldots , 
\\
& R_{\mu\nu\al\be} R^{\mu\nu\al\be}   \underset{u \to 0}{=}  \tfrac{3 h_0^2 (2 p^2-4 p q+3 q^2) (6 p^2-16 p q+11 q^2)}{4 q^4} 
 u^{2 (p-2 q)} + \ldots
\nonumber
\end{split}
\eeq
are singular at $u = 0$ because $p-2 q < 0$ and the above prefactors cannot vanish on the relevant domain of $p$ and $q$. The behaviour in terms of the coordinate $r$, as $r \to r_0$, follows from $u = \Delta^{1/q}$ [for instance, $R \sim r^{-7/10}$ and $R_{\mu\nu\al\be} R^{\mu\nu\al\be} \sim r^{-7/5}$ for the example~\eqref{example}]. In Schwarzschild coordinates and in terms of the parameter 
$s_* <0$, we have
\beq
\nonumber
R \underset{\bar{r} \to 0}{\sim} \bar{r}^{s_*-2} , \qquad  R_{\mu\nu\al\be} R^{\mu\nu\al\be} \underset{\bar{r} \to 0}{\sim} \bar{r}^{2(s_*-2)}.
\eeq
The singularity of the Kretschmann invariant can be weaker or stronger than that of the other solutions of quadratic gravity.\footnote{Namely, $R_{\mu\nu\al\be} R^{\mu\nu\al\be} \sim \bar{r}^{-6}$ for the $(-1,-1)_0$ solutions and $R_{\mu\nu\al\be} R^{\mu\nu\al\be} \sim \bar{r}^{-8}$ for the $(-2,2)_0$ ones~\cite{Stelle78}.} Specifically, for models with $\omega > 25$ we have $-1<s_*<0$ and this family of solutions has the mildest singularity. If $25/4 < \omega < 25$, then $-2<s_*<-1$ and the Kretschmann scalar diverges faster than the solutions $(-1,-1)_0$, but slower than the $(-2,2)_0$ ones. Finally, this solution family has the strongest singularity if $1 < \omega < 25/4$, for which $s_*<-2$. For the special cases of $\omega=25$ and $\omega=25/4$ the solutions in the family $(s_*, 2-3s_*)_0$ diverge similarly to those in the families $(-1,-1)_0$ and $(-2,2)_0$, respectively.

To summarise, in this section we showed that all quadratic gravity models such that $\tau_+(\omega)$ [defined in Eq.~\eqref{TauInv}] is a rational number 
admit a family of solutions with indicial structure $\lbrace  \sigma ,\tau \rbrace =  \left\lbrace 3/2 - \tau_+ , \tau_+ \right\rbrace $ in coordinates~\eqref{metric}. These models actually form a dense set in the space of models with $\omega>1$. In fact, in view of Eq.~\eqref{def2}, consider the function
\beq
{\bar{\omega}(\bar{\tau}) = \frac{(2 \bar{\tau}^2 - 6 \bar{\tau} + 5)^2}{4 (\bar{\tau}-1)^4}, \qquad \bar{\tau} \in (1,3/2)}.
\eeq
Then $\bar{\omega}$ is monotonic, continuous and its image is $(1,\infty)$. Since $(1,3/2)\cap\mathbb{Q}$ is dense in $(1,3/2)$, the continuous image $\bar{\omega}((1,3/2)\cap\mathbb{Q})$ is also dense in $(1,\infty)$. Therefore, for any real $\bar{\omega}_1>1$ and $\epsilon>0$, there exists a rational $\tau_1\in (1,3/2)\cap\mathbb{Q}$ such that ${|\bar{\omega}(\tau_1)-\bar{\omega}_1| < \epsilon}$. The density of these models may suggest that such solutions also exist for irrational values of $\tau$ and in all the models with $\omega>1$, but probably they take a more complicated, beyond-Frobenius form in any of the coordinates considered here.

\section{The family $(s_*, 2-3s_*)_\infty$}
\label{Sec3}

In addition to~\eqref{def2}, the indicial equations in modified coordinates~\eqref{metric} admit a related solution, the only difference being that now $3/2 < \tau <2$. This suggests the existence of a new family of solutions with $\lbrace  \sigma ,\tau \rbrace =  \left\lbrace  3/2 - \tau_-(\omega) , \tau_-(\omega) \right\rbrace $ for models with $1/4 < \omega < 1$, where $\tau_-(\omega)$ is uniquely defined by [\textit{cf.} Eq.~\eqref{TauInv}]
\beq
\label{TauInvInf}
\tau_- (\omega) = \frac{1}{2} \left( 2 - \frac{1}{\sqrt{\omega }-1} -  
\sqrt{\frac{2 \sqrt{\omega }-1}{\big(\sqrt{\omega }-1\big)^2}}\right) , \qquad  \frac{1}{4} < \omega < 1 .
\eeq
The analysis carried out in the previous section can be repeated for this family noting that the modification in the range of $\tau$ now yields $p/2 < q< 2 p /3$ for $\tau\in\mathbb{Q}$. This does not affect, however, the formal change to coordinates~\eqref{metric-U}, the structure of the field equations expanded order by order~\eqref{E_N}, nor the count of free parameters (assuming $\mathfrak{d} \neq 0$). (Numerical searches for roots of the Diophantine equation $\mathfrak{d} = 0$ on this new parameter domain did not return any solution for $p$ smaller than 5000.) Finally, the argument used at the end of Sec.~\ref{Sec2} can be applied, \textit{mutatis mutandis}, to prove that the set of models that admit the solutions $\left\lbrace  3/2 - \tau_-(\omega) , \tau_-(\omega) \right\rbrace $ also forms a dense set, albeit in the considerably smaller set of models with $\omega\in (1/4,1)$.

Regarding the description in standard Schwarzschild coordinates,
now the range of $\tau_-(\omega)$ results in $\sigma = 3/2 - \tau_-  <0$. Therefore, this family of solutions represents asymptotic expansions 
[indeed, $r \to r_0$ corresponds to the limit $\bar{r} \to \infty$, see~\eqref{Fr}], 
which we denote by $(s_*, 2-3s_*)_\infty$ where $s_*$ is  given by~\eqref{star_def}, now restricted to $s_* \in (2,\infty)$. 
(It is worth mentioning that~\eqref{st1} can also be used to relate solutions $\left\lbrace  \sigma , \tau \right\rbrace$ with $\sigma<0$ and $(s,t)_\infty$,  see~\cite{Giacchini:2025gzw} for details.)
For integer values of $s_*$, the functions $A$ and $B$ assume the form of Frobenius series in $\bar{r}^{-1}$; otherwise, they may involve non-integer steps. As an explicit example, models with $\omega = 25/81$ admit solutions $\left\lbrace  -1/4 , 7/4 \right\rbrace$,
which via~\eqref{st1} yields $(3,-7)_\infty$:
\beq
\nonumber
\begin{split}
A(\bar{r}) & = \bar{r}^3 \left( a_0 - \tfrac{308 \gamma }{573 \beta } \bar{r}^{-1} + O (\bar{r}^{-2}) \right)   , 
\\ 
B(\bar{r}) & = \frac{b_0}{\bar{r}^7} \left( 1 + \tfrac{39 \gamma }{191 a_0 \beta } \bar{r}^{-1} + O (\bar{r}^{-2}) \right) .
\end{split}
\eeq

The solutions in this family have diverging curvatures 
as $\bar{r}\to\infty$, for instance, $R \sim \bar{r}^{s_*-2}$ and $R_{\mu\nu\al\be} R^{\mu\nu\al\be} \sim \bar{r}^{2(s_*-2)}$. 
The point ${\bar{r}=\infty}$ can be interpreted as a true spacetime singularity, since it can be reached at a finite value of affine parameter by causal geodesics. This can be seen by considering radial null geodesics $t(\lambda)$, $\bar{r}(\lambda)$, with the tangent vector ${k^{\mu}=(\dot{t},\dot{\bar{r}},0,0)}$. Putting together the null condition, ${k^{\mu}k_{\mu}=-B\dot{t}^2+A^{-1}\dot{\bar{r}}^2=0}$, and the energy conservation, $-E={k_{\mu}(\partial/\partial t)^{\mu}=-B\dot{t}}$, we arrive at $\dot{\bar{r}}=\pm E/\sqrt{B/A}$. This upon inverting and integrating for the affine parameter leads to ${\Delta\lambda=E^{-1}\int^{\infty}\sqrt{B/A}\rd\bar{r}}$. Since ${B\sim \bar{r}^{2-3s_*}}$ and ${A\sim \bar{r}^{s_*}}$ this integral converges if and only if ${2-4s_*<-2}$, which is certainly satisfied for ${s_*>2}$. Hence, as the area of the 2-spheres with 
${\bar{r}, t = \textrm{const.}}$ grows, the spacetime becomes so 
distorted that it terminates at a finite affine distance, 
corresponding to a genuine singular boundary rather than any 
conformal infinity.
This singularity has a null character, inasmuch as $B(\bar{r}) \to 0$ for $\bar{r}\to\infty$.

\section{New solutions around $\bar{r}_0\neq 0$.}
\label{Sec3b}

We also report the finding of new solutions corresponding to non-Frobenius expansions around a point $\bar{r} = \bar{r}_0 \neq 0$ in Schwarzschild coordinates. In modified coordinates~\eqref{metric}, however, they appear as subcases of the large families $\left\lbrace \sigma,\tau \right\rbrace=\left\lbrace 0,0 \right\rbrace$ and $\left\lbrace 0,1 \right\rbrace$, which contain several well-known solutions. 
For generic solutions with $\sigma = 0$, instead of~\eqref{st1}, the leading exponents $\left\lbrace \sigma,\tau \right\rbrace$ and $(s,t)$ relate through
\beq
\label{st3}
s =  \frac{\tau}{N} + \frac{2(N-1)}{N}   , \qquad t = \frac{\tau}{N} ,
\eeq
where $N\in\mathbb{N}$ is such that
$f_N\neq 0$ is the first nonzero coefficient after $f_0$~\cite{Giacchini:2025gzw}.

General solutions in the $\left\lbrace 0,0 \right\rbrace$ family have the form of Frobenius series [like in Eq.~\eqref{Series-FH} with $q=1$ and $\sigma=\tau=0$] with a total of 8 free parameters: $r_0$ and 7 among $f_{0,\ldots,3}$ and $h_{0,\ldots,3}$, the constraint being given by the component $rr$ of the field equations at the lowest order, $(\mathcal{E}^r{}_r)_0 = 0$. (Two parameters can be fixed using the gauge freedom of the metric~\eqref{metric}, so there are 6 physical parameters.) 
From~\eqref{Fr} and~\eqref{st3},
solutions with $f_1 \neq 0$ describe expansions around a generic point $r=r_0$, equivalent to the family $(0,0)_{\bar{r}_0}$ in Schwarzschild coordinates~\cite{Stelle15PRD}. However, $r_0$ is interpreted as a wormhole throat~\cite{Visser:1995cc} if $f_{1}=0$ (see~\cite{Giacchini:2025gzw} for further details; the situation here is also similar to what happens with these solutions in Kundt coordinates, as described in~\cite{Podolsky:2019gro}). For example, solutions with $f_1=0$ but $f_2\neq 0$ are  equivalent to the non-Frobenius solutions $(1,0)_{\bar{r}_0, 1/2}$ in Schwarzschild coordinates~\cite{Stelle15PRD}, and $f_3=h_{1,3}=0$ yields the Frobenius solutions $(1,0)_{\bar{r}_0}$~\cite{Stelle15PRD}. Also, if $f_{1,2}=0$ we obtain the family $(4/3,0)_{\bar{r}_0, 1/3}$. The solutions in this latter family such that $R=0$ were first identified in~\cite{Podolsky:2019gro}; our analysis shows the existence of an extension of these solutions with 2 additional parameters so that $R\neq 0$.

New families of solutions occur if $f_{1,2,3}=0$ and $\omega \neq 1$ (in this case, the constraint $(\mathcal{E}^r{}_r)_0 = 0$ cannot be satisfied if $\omega = 1$). A generic solution of this type is characterised by 3 physical free parameters and, if $f_4\neq 0$, corresponds to a non-Frobenius solution $\left( 3/2,0\right)_{\bar{r}_0, 1/4}$ in Schwarzschild coordinates.
Similarly, the even version of this solution (\textit{i.e.}, with the additional requirement $h_{1,3} =0$) corresponds to a solution $\left( 3/2,0\right)_{\bar{r}_0, 1/2}$ 
with a single free physical parameter. 
Alternatively, it is possible that the parameters $f_0$, $h_{1,2}$ conspire so that $f_4=0$ or even $f_{4,5}=0$, resulting in the structures $(8/5,0)_{\bar{r}_0,1/5}$ and
$(5/3,0)_{\bar{r}_0,1/6}$ in Schwarzschild coordinates. However, due to the way $f_{4,5}$ depend on the lower-order coefficients, these solutions do not occur for all values of couplings and coefficients.
Specifically, $f_4=0$ requires
\beq
\label{condf2=0}
\frac{\beta  [\gamma ^2 (\omega-1)f_0^4 -24  \gamma \beta  (\omega+2) f_0^2 +144 \beta ^2 (\omega-1)]}{\omega-1} \geqslant 0,
\eeq
which might constrain the possible values of $f_0$. For $f_{4,5}=0$,
\beq
\label{condf2=f3=0}
\omega > 1 \,
\text{(if  $\gamma\beta > 0$),} 
\,\, \quad 
\text{ or } \quad -1/2 \leqslant \omega < 1 
\,\,
\text{(if  $\gamma\beta < 0$).} 
\eeq
To our knowledge, these solutions were not identified before.

A similar situation occurs with subcases of the family $\left\lbrace 0,1 \right\rbrace$ [see Eq.~\eqref{Series-FH} with $q=1$, $\sigma=0$ and $\tau=1$], which has the free parameters $\lbrace f_{0,1},h_{0,1},r_0\rbrace$ and contains expansions around horizons. While a generic solution $\left\lbrace 0,1 \right\rbrace_{f_1\neq 0}$ represents a Schwarzschild-like horizon of type $(1,1)_{\bar{r}_0}$~\cite{Stelle15PRL,Stelle15PRD}, the case $\left\lbrace 0,1 \right\rbrace_{f_1=0,f_2\neq 0}$ yields the non-Frobenius family $(3/2,1/2)_{\bar{r}_0,1/2}$~\cite{Stelle15PRD}.
In addition, if $\omega \neq 1$ and~\eqref{condf2=0} is satisfied, it is possible to tune $f_0$ and $h_1$ so that $f_2=0$. If~\eqref{condf2=f3=0} holds, then one can have $f_{2,3}=0$. These sub-families correspond in Schwarzschild coordinates to the structures $(5/3,1/3)_{\bar{r}_0,1/3}$ and
$(7/4,1/4)_{\bar{r}_0,1/4}$, respectively.

\begin{table*}[t]
    \centering
    \begin{tabular}{|c|c|c|c|c|c|}
            \hline
             \multicolumn{2}{|c|}{Solution family}  & \multirow{2}{*}{Parameters} & Number of free & \multirow{2}{*}{Interpretation} & \multirow{2}{*}{Remarks} \\ 
            $(s,t)$  & $\lbrace \si,\ta \rbrace$   & 	       & parameters &  & \\ 
            \hline
            \hline
            $(0,0)_0$ & $\lbrace 1,0 \rbrace$ & $f_0$, $f_2$, $h_2$, $r_0$ & $ 4 \to 2 $ & regular core & \\ 
            \hline
            \hline
            $(-1,-1)_0$ & $\lbrace 1,-1 \rbrace$ & $f_0$, $f_3$, $h_0$, $h_3$, $r_0$ & $ 5 \to 3 $ & Schwarzschild-like core & \\ 
            \hline
            \hline
            $(-2,2)_0$ & $\left\lbrace   \tfrac{1}{3},\tfrac{2}{3} \right\rbrace $  (nF) & $f_0$, $f_1$, $f_2$, $f_3$, $h_0$, $h_3$, $r_0$ & $ 7 \to 5 $ & 2-2-hole core & \\ 
            \hline
            \hline
            \multirow{2}{*}{$(s_*, 2-3s_*)_0$ (nF)} &  \multirow{2}{*}{$\left\lbrace   \tfrac{3}{2} - \tau_+, \tau_+ \right\rbrace $  (nF)}  & \multirow{2}{*}{$f_0$, $h_0$, $r_0$} & \multirow{2}{*}{$ 3 \to  1$} & \multirow{2}{*}{singular core} & $\omega>1$, $\,\, \tau_+ \in \mathbb{Q}$, \\
            & & & & & $\tau_+$ in Eq.~\eqref{TauInv} \\ 
            \hline
            \hline
            $(0,0)_{\bar{r}_0}$ & $\lbrace 0,0 \rbrace$ & ($f_0$, $f_1$, $f_2$, $f_3$, $h_0$, $h_1$, $h_2$, $h_3$), $r_0$ & $8 \to 6$ & generic point & \\ 
            \hline  
            $(1,0)_{\bar{r}_0, 1/2}$  (nF) & $\lbrace 0,0 \rbrace_{f_1=0, \,f_2\neq 0}$ & ($f_0$, $f_2$, $f_3$, $h_0$, $h_1$, $h_2$, $h_3$), $r_0$ & $7 \to 5$ & non-symmetric throat & \\ 
            \hline  
            $(1,0)_{\bar{r}_0}$ & $\lbrace 0,0 \rbrace_{f_{1,3}=h_{1,3}=0, \,f_2\neq 0}$ & ($f_0$, $f_2$, $h_0$, $h_2$), $r_0$ & $4 \to 2$ & symmetric throat & \\ 
            \hline  
            $\left( \tfrac{4}{3},0\right) _{\bar{r}_0, 1/3}$  (nF) & $\lbrace 0,0 \rbrace_{f_{1,2}=0, \,f_3\neq 0}$ & ($f_0$, $f_3$, $h_0$, $h_1$, $h_2$, $h_3$), $r_0$ & $6 \to 4$ & non-symmetric throat & \\ 
            \hline  
            $\left( \tfrac{3}{2},0\right) _{\bar{r}_0, 1/4}$  (nF) & $\lbrace 0,0 \rbrace_{f_{1,2,3}=0, \,f_4\neq 0}$ & ($f_0$, $h_0$, $h_1$, $h_2$, $h_3$), $r_0$ & $5 \to 3$ & non-symmetric throat & $\omega \neq 1$ \\ 
            \hline  
            $\left( \tfrac{3}{2},0\right) _{\bar{r}_0, 1/2}$  (nF) & $\lbrace 0,0 \rbrace_{f_{1,2,3}=h_{1,3}=0, \,f_4\neq 0}$ & ($f_0$, $h_0$, $h_2$), $r_0$ & $3 \to 1$ & symmetric throat & $\omega \neq 1$ \\ 
            \hline  
            $\left( \tfrac{8}{5},0\right) _{\bar{r}_0, 1/5}$  (nF)  & $\lbrace 0,0 \rbrace_{f_{1,2,3,4}=0, \,f_5\neq 0}$  & ($f_0$, $h_0$, $h_1$, $h_2$), $r_0$  &  $4 \to 2$  & non-symmetric throat  &  $\omega \neq 1$ and Eq.~\eqref{condf2=0} \\ 
            \hline  
            $\left( \tfrac{5}{3},0\right) _{\bar{r}_0, 1/6}$  (nF)  &  $\lbrace 0,0 \rbrace_{f_{1,2,3,4,5}=0, \,f_6\neq 0}$  &  $h_0$, $h_1$, $r_0$  &  $3 \to 1$  &  non-symmetric throat  &  $\omega \neq 1$, see Eq.~\eqref{condf2=f3=0} \\ 
            \hline
            \hline
            $(1,1)_{\bar{r}_0}$ & $\lbrace 0,1 \rbrace$ & $f_0$, $f_1$, $h_0$, $h_1$, $r_0$ & $5 \to 3$ & Schwarzschild-like horizon & \\ 
            \hline 
            $\left( \tfrac{3}{2},\tfrac{1}{2} \right)_{\bar{r}_0,1/2}$  (nF) & $\lbrace 0,1 \rbrace_{f_1=0, \, f_2 \neq 0}$ & $f_0$, $h_0$, $h_1$, $r_0$ & $4 \to 2$ &  horizon & \\ 
            \hline 
            $\left( \tfrac{5}{3},\tfrac{1}{3} \right)_{\bar{r}_0,1/3}$  (nF)  &  $\lbrace 0,1 \rbrace_{f_{1,2}=0, \, f_3 \neq 0}$  &  $f_0$, $h_0$, $r_0$  &  $3 \to 1$  &   horizon   &  $\omega\neq 1$ and Eq.~\eqref{condf2=0}  
            \\ 
            \hline 
            $\left( \tfrac{7}{4},\tfrac{1}{4} \right)_{\bar{r}_0,1/4}$  (nF) & $\lbrace 0,1 \rbrace_{f_{1,2,3}=0, \, f_4 \neq 0}$  &  $h_0$, $r_0$  &  $2 \to 0$  & horizon & $\omega\neq 1$ and Eq.~\eqref{condf2=f3=0}
            \\ 
            \hline
            \hline 
            \multirow{2}{*}{$(s_*, 2-3s_*)_\infty$ (nF)} & \multirow{2}{*}{$\left\lbrace   \tfrac{3}{2} - \tau_-, \tau_- \right\rbrace $ (nF)} & \multirow{2}{*}{$f_0$, $h_0$, $r_0$} & \multirow{2}{*}{$3 \to  1$}  & \multirow{2}{*}{singular boundary} & $1/4 < \omega <1$, $\,\,\tau_- \in \mathbb{Q}$, \\
            & & & & & $\tau_-$ in Eq.~\eqref{TauInvInf} \\ 
            \hline
            \hline 
            $(0,0)_\infty$ & $\lbrace 1,0 \rbrace^\infty$ & $f_0$, $f_1$, $h_1$ & $ 3 \to 1$ & Schwarzschild &  \\ 
            \hline
        \end{tabular}  
        \caption{\small Summary of solutions in quadratic gravity models with $\alpha,\beta,\gamma\neq 0$. 
        The first two columns indicate the indicial structure in standard spherically symmetric coordinates~\eqref{metric-Standard} and the family or sub-family in modified coordinates~\eqref{metric}; solutions not necessarily of Frobenius form are indicated by ``nF''. The third column lists the free parameters of the solution in the form~\eqref{Series-FH}; parentheses indicate a constraint between them, so the total number of free parameters is one less. The count of free parameters is listed in the fourth column, with the arrow indicating the reduction to the physical ones after taking into account the residual gauge freedom of the metric~\eqref{metric}. The last columns provide the interpretation and remarks about the new solutions.}
\label{Tab1}
\end{table*}

\section{Summary of solutions in modified Schwarzschild coordinates}
\label{Sec4}

In the course of this work, we carried out a complete classification of static and spherically symmetric vacuum solutions in general quadratic gravity that admit a Frobenius series expansion when written in the form~\eqref{metric}. This analysis serves the purpose of identifying how the previously known solutions are expressed in these 
coordinates; in addition, it led to the discovery of new families of solutions that do not have Frobenius form in standard coordinates~\eqref{metric-Standard}, as described in 
the previous sections.
We omit the technical details in this summary, as the majority of families are already known and different aspects of the solutions were studied, \textit{e.g.}, in~\cite{Stelle78,Stelle15PRL,Stelle15PRD,Holdom:2002,Podolsky:2018pfe,Podolsky:2019gro,Holdom:2016nek,Holdom:2022zzo,Goldstein:2017rxn,Bonanno:2019rsq,Salvio:2019llz,Daas:2022iid,Bonanno:2022ibv,Silveravalle:2022wij}.

The strategy followed was the same applied in~\cite{Giacchini:2025gzw} in the context of six-derivative gravity, and we refer the interested reader to that work for further details. We considered two types of solutions: for expansions around a point $r=r_0$, we assume that the functions $F$ and $H$ in~\eqref{metric} have the leading behaviour $F(r) \sim \Delta^\sigma$ and $H(r) \sim \Delta^\tau$, followed by higher powers of $\Delta=r-r_0$, whereas for asymptotic expansions ($r\to\infty$) we assume series in powers of $r^{-1}$ with the leading terms $F(r) \sim r^\sigma$ and $H(r) \sim r^\tau$.
The former solutions are denoted by the indicial structure $\lbrace \sigma,\tau \rbrace$, and the latter by $\lbrace \sigma,\tau \rbrace^\infty$.  Like in a standard Frobenius analysis, the field equations at leading order fixes the admissible indicial structures. For a generic quadratic gravity model with $\alpha,\beta,\gamma \neq 0$ the only possible solutions are
\beq
\label{SolIndEq}
\begin{split}
& \left\lbrace 0,0 \right\rbrace , \quad 
\left\lbrace 0,1 \right\rbrace , \quad 
\left\lbrace 1,0 \right\rbrace , \quad 
\left\lbrace 1,-1 \right\rbrace , \quad 
\left\lbrace \tfrac{1}{3},\tfrac{2}{3}  \right\rbrace , 
\\
& \left\lbrace \tfrac{3}{2} - \tau_+(\omega) , \tau_+(\omega) \right\rbrace , \quad
\left\lbrace \tfrac{3}{2} - \tau_-(\omega) , \tau_-(\omega) \right\rbrace , \quad
\left\lbrace 1,0 \right\rbrace^\infty ,
\end{split}
\eeq
with $\tau_+(\omega)$ defined in~\eqref{TauInv} and $\tau_-(\omega)$ in~\eqref{TauInvInf}. 

Solving the field equations to higher orders, one realizes that the solutions $\left\lbrace 1/3, 2/3  \right\rbrace$ cannot have a Frobenius expansion in $\Delta$; instead, they are of the form~\eqref{Series-FH} with $q=3$, \textit{i.e.},  with increments of $\Delta^{1/3}$. These solutions correspond to the family $(-2,2)_0$ in Schwarzschild coordinates, as can be checked by comparing the indicial structures [see Eq.~\eqref{st1}] and the count of free parameters in~\cite{Holdom:2002,Stelle15PRD}. The solutions $\left\lbrace 3/2 - \tau_\pm(\omega) , \tau_\pm(\omega) \right\rbrace$ generally are not of Frobenius form either, as discussed above.
It is unrealistic to explore all the forms of non-Frobenius solutions but, \emph{out of necessity}, we investigated this possibility only for these three families.

The other structures in~\eqref{SolIndEq} admit power series solutions. 
For each of them, we proved that there exist recursive relations for the series coefficients, which yields a precise determination of the number of free parameters. To this end, we used similar arguments as those 
involving Eq.~\eqref{E_N}
and of Ref.~\cite{Giacchini:2025gzw}. Our results are summarised in Table~\ref{Tab1}, together with the interpretation of the solutions and the description in standard spherically symmetric coordinates. The count of physical free parameters of the solutions that were previously known coincides with~\cite{Stelle15PRD}.


\section{Closing remarks}
\label{Sec5}

In higher-derivative gravity, although it is natural to search for solutions expandable as series
with exponents that do not depend on the parameters of the model, it is just as reasonable to conceive that  more complicated solutions exist. Here we explored this possibility, proving the existence of sets of models that are dense in certain portions of the parameter space and that admit solutions with exponents that depend on the ratio of the quadratic couplings. Our proof assumed a specific form for the solution, as series whose exponents increase by rational numbers. It might be possible that such solutions can be extended to irrational increments and, therefore, to a set of models with non-zero measure in the space of models.

Another possible extension of these solutions could be in the form of series with smaller increments than those considered here. Actually, any family in~\eqref{SolIndEq} could in principle be part of a larger family with smaller steps.
Although a full classification of non-Frobenius solutions is unrealistic, our results motivate further study of solutions with coupling-dependent exponents.

Such solutions might play an important role in general models with more than four derivatives, as the large number of coupling constants may restrict the solution families with coupling-independent exponents. For instance, the only solutions of this type describing the origin $\bar{r}=0$ in six-derivative gravity have the structure $(0,0)_0$~\cite{Giacchini:2025gzw,Giacchini:2024exc}, and there is evidence that the situation is similar in higher-order models~\cite{Holdom:2002}. At the same time, specific models are known to admit other solutions~\cite{Giacchini:2024exc}. A more detailed discussion on the case of six-derivative gravity will be presented in a separate publication.

Finally, evoking the argument of~\cite{Stelle15PRD} that relates asymptotic flatness, presence of horizons and the vanishing of the Ricci scalar, one concludes that none of the new solutions presented here can be both asymptotically flat and have a horizon, since all have $R \neq 0$. For this same reason, they are not solutions of Einstein--Weyl gravity and could not have been found in~\cite{Podolsky:2019gro}.

\begin{acknowledgements}
\noindent
We thank Otakar Sv\'itek for pointing out the null character of the singularity of the solutions $(s_*,2-3s_*)_\infty$. 
Both authors acknowledge financial support by the Primus grant PRIMUS/23/SCI/005 from Charles University and the support from the Charles University Research Center Grant No. UNCE24/SCI/016.
The calculations in this work were performed using \textit{Mathematica} and the package \textsc{xAct}.
\end{acknowledgements}

\appendix

\begin{widetext}

\section{Explicit formulas for the coefficients of the linear system~\eqref{E_N}}
\label{SM}

The coefficients $\mathfrak{f}_{1,2}$ and $\mathfrak{h}_{1,2}$ of the system~\eqref{E_N} read
%
{\allowdisplaybreaks
\begin{eqnarray*}
\mathfrak{f}_1(p,q,N) & = & 
4 N^3 (2 p^2-4 p q+3 q^2) (6 p^2-16 p q+11 q^2)
-6 N^2 (p-q) (2 p^2-4 p q+3 q^2) (6 p^2-16 p q+11 q^2)
\nonumber
\\
&&
-2 N (p-q)^2 (36 p^4-268 p^3 q+700 p^2 q^2-784 p q^3+327 q^4)
+36 (p-q)^3 (p-2 q)^2 (2 p^2-6 p q+5 q^2)
,
\nonumber
\\
\mathfrak{f}_2(p,q,N) & = & 
2 \big[N^2 (4 p^4-16 p^3 q+16 p^2 q^2+4 p q^3-9 q^4)
+N (p-q) (4 p^4-12 p^3 q+16 p^2 q^2-16 p q^3+9 q^4)
\nonumber
\\
&&
-6 p (p-q)^2 (p-2 q) (2 p^2-6 p q+5 q^2)\big],
\nonumber
\\
\mathfrak{h}_1(p,q,N) & = & 
(2 p-3 q) \big[2 N^3 q (4 p^2-10 p q+7 q^2)
-3 N^2 (p-q) (4 p^3-16 p^2 q+22 p q^2-11 q^3)
\nonumber
\\
&&
+2 N (p-q)^2 (18 p^3-70 p^2 q+100 p q^2-55 q^3)
+18 (p-q)^3 (p-2 q) (2 p^2-6 p q+5 q^2)\big]
,
\nonumber
\\
\mathfrak{h}_2(p,q,N) & = & (2 p-3 q)
\big[N^2 (4 p^3-16 p^2 q+22 p q^2-11 q^3)
+2 N (p-q) (2 p^3-6 p^2 q+8 p q^2-5 q^3)
\nonumber
\\
&&
-6 (p-q)^2 (p-2 q) (2 p^2-6 p q+5 q^2)\big].
\end{eqnarray*}
%
}
\end{widetext}

\end{document}